\documentclass{moriond}

\def\Journal#1#2#3#4{{#1} {\bf #2}, #3 (#4)}

\def\be{\begin{equation}}
\def\ee{\end{equation}}
\def\bea{\begin{eqnarray}}
\def\eea{\end{eqnarray}}

\newcommand{\Photo}{}

\begin{document}
\vspace*{4cm}
\title{High angular resolution SZ observations with NIKA and NIKA2}

\author{B.~Comis$^1$,
R.~Adam$^{1,2}$,
P.~Ade$^3$,
P.~Andr\'e$^4$,
M.~Arnaud$^4$,
I.~Bartalucci$^4$,
A.~Beelen$^5$,
A.~Beno\^it$^6$,
A.~Bideaud$^3$,
N.~Billot$^7$,
O.~Bourrion$^1$,
M.~Calvo$^6$,
A.~Catalano$^1$,
G.~Coiffard$^9$,
F.-X.~D\'esert$^{10}$,
S.~Doyle$^3$,
J.~Goupy$^6$,
C.~Kramer$^7$,
G.~Lagache$^8$,
S.~Leclercq$^9$,
J.~F.~Mac\'ias-P\'erez$^1$,
P.~Mauskopf$^{3,11}$,
F.~Mayet$^1$,
A.~Monfardini$^6$,
F.~Pajot$^5$,
E.~Pascale$^3$,
L.~Perotto$^1$,
E.~Pointecouteau$^{12}$,
G.~Pisano$^3$,
N.~Ponthieu$^{10}$,
G.~W.~Pratt$^4$,
V.~Rev\'eret$^4$,
A.~Ritacco$^1$,
L.~Rodriguez$^4$,
C.~Romero$^9$,
F.~Ruppin$^1$,
G.~Savini$^{13}$,
K.~Schuster$^9$,
A.~Sievers$^7$,
S.~Triqueneaux$^6$,
C.~Tucker$^3$,
R.~Zylka$^9$}

\address{$^1$ Laboratoire de Physique Subatomique et de Cosmologie, Universit\'e Grenoble Alpes, CNRS/IN2P3, 53, avenue des Martyrs, Grenoble, France\\
 $^2$Laboratoire Lagrange, Universit\'e C\^ote d'Azur, Observatoire de la C\^ote d'Azur, CNRS, Blvd de l'Observatoire, CS 34229, 06304 Nice cedex 4, France\\
 $^3$Astronomy Instrumentation Group, University of Cardiff, UK \\
 $^4$Laboratoire AIM, CEA/IRFU, CNRS/INSU, Universit\'e Paris Diderot, CEA-Saclay, 91191 Gif-Sur-Yvette, France\\
 $^5$Institut d'Astrophysique Spatiale (IAS), CNRS and Universit\'e Paris Sud, Orsay, France\\
 $^6$Institut N\'eel, CNRS and Universit\'e Grenoble Alpes, France\\
 $^7$Institut de RadioAstronomie Millim\'etrique (IRAM), Granada, Spain\\
 $^8$Aix Marseille Universit\'e, CNRS, LAM (Laboratoire d'Astrophysique de Marseille) UMR 7326, 13388, Marseille, France\\
 $^9$Institut de RadioAstronomie Millim\'etrique (IRAM), Grenoble, France\\
 $^{10}$Institut de Plan\'etologie et d'Astrophysique de Grenoble (IPAG), CNRS and Universit\'e de Grenoble, France\\
 $^{11}$School of Earth and Space Exploration and Department of Physics, Arizona State University, Tempe, AZ 85287\\
 $^{12}$Universit\'e de Toulouse, UPS-OMP, Institut de Recherche en Astrophysique et Plan\'etologie (IRAP), Toulouse, France\\
 $^{13}$University College London, Department of Physics and Astronomy, Gower Street, London WC1E 6BT, UK}

\maketitle\abstracts{NIKA2 (New IRAM KID Arrays) is a dual band (150 and 260 GHz) imaging camera based on Kinetic Inductance Detectors (KIDs) and designed to work at the IRAM 30 m telescope (Pico Veleta, Spain). Built on the experience of the NIKA prototype, NIKA2 has been installed at the 30 m focal plane in October 2015 and the commissioning phase is now ongoing. Through the thermal Sunyaev-Zeldovich (tSZ) effect, NIKA2 will image the ionized gas residing in clusters of galaxies with a resolution of 12 and 18 arcsec FWHM (at 150 and 260 GHz, respectively). We report on the recent tSZ measurements with the NIKA camera and discuss the future objectives for the NIKA2 SZ large Program, 300h of observation dedicated to SZ science. With this program we intend to perform a high angular resolution follow-up of a cosmologically-representative sample of clusters belonging to SZ catalogues, with redshift $\geq$ 0.5. The main output of the program will be the study of the redshift evolution of the cluster pressure profile as well as that of the scaling laws relating the cluster global properties.}

\section{Introduction}
Clusters of galaxies are the largest gravitationally bound structures observable in our Universe. They represent the last step of the hierarchical structure formation process and so their number and distribution as a function of mass and redshift can provide an excellent tool for cosmology. Being complementary to purely geometric probes and observables of the primordial Universe, clusters can provide tight constraints on the matter content and distribution within our Universe but also on the origin of its accelerated expansion. 
However, to do cosmology with clusters, we must be able to translate the direct observables, used to detect and count them, into precise mass estimates. The accuracy and the precision of the total mass estimates will in fact directly impact the robustness of the derived cosmological constraints.

Clusters of galaxies are mainly ($\sim$ 80\%) constituted of dark matter, while most of the baryons are in the form of a hot ionized gas, the so-called intra-cluster medium (ICM). Then, cluster total masses can be measured through their gravitational effect, through the gravitational lensing, or by using their baryonic components as a tracer of the total mass distribution. In particular, the observables related to their main baryonic component, the ICM, have been proven to be a good proxy of the cluster total mass. Given its temperature (10$^6$~-~10$^8$~K) and density (10$^{-4}$~-~10$^{-2}$~cm$^{-3}$) this gas is responsible for the Bremsstrahlung X-ray emission, but it also produces a secondary anisotropy of the Cosmic Microwave Background (CMB), which is the thermal Sunyaev-Zel'dovich (tSZ) effect.
The tSZ effect is a distortion of the black body CMB spectrum produced by the inverse Compton interaction of CMB photons with the hot electrons of the ICM \cite{{SZ1},{SZ2},{SZ3}}. This interaction produces a unique spectral signature, with a decreased and increased CMB intensity at frequencies respectively lower and higher than 217~GHz. While the frequency behavior is, at first order, specific of the effect, the amplitude of the signal, given by the Comptonization parameter $y$, is instead related to the cluster thermal energy content, and then to its mass, being proportional to the integral of the gas pressure along the line of sight ($y = \sigma_{\mathrm{T}}/m_{\mathrm{e}} c^2 \int P_{\mathrm{e}} dl$, with $m_e$ the electron mass, $c$ the light speed, $\sigma_T$ the electron Thomson scattering cross-section). A third interesting point about SZ is that, since we are dealing with a spectral deformation, it is not affected by the dilution with redshift. Making it particularly well adapted to detect clusters at higher $z$ where their number and distribution is the most sensitive to the underlying cosmology.

During the last decade, the first cluster derived cosmological constraints were obtained, initially mainly using X-ray data and more recently also with tSZ. In fact, during the last years, thanks to survey dedicated experiments like Planck \cite{PlanckXXIX}, ACT \cite{Hasselfield} and SPT \cite{Bleem}, we rapidly went from the first SZ-discovered clusters \cite{Staniszewski2009} to SZ-selected cluster catalogs containing more than thousand of sources, enabling to estimate cosmological parameters ($\sigma_8$, $\Omega_m$). However these results have shown tension with those derived from primary CMB anisotropies. Anyway, in order to interpret further this tension and an eventual cosmological explanation for it, we must control any possible systematic. On CMB side the tension has been reduced by the latest Planck constraints on the optical depth of reionization \cite{PlanckXIII}, while on the cluster side we need to further explore the presence and the amplitude of astrophysical systematics that might be introduced by an incomplete knowledge of the details of cluster physics and the different assumption made. Indeed, at present, tSZ derived cosmological constraints lay on several assumptions. These are the assumption of hydrostatic equilibrium between the gas and the overall cluster potential well and the self-similarity of the cluster population. Furthermore we also extrapolate up to intermediate redshift ($\sim$~0.5) and beyond the calibration of cluster properties that have been obtained observationally in the local universe ($\sim$~0.2). In order to make clusters a tool for precision cosmology we still need precise calibrations of the mass-observable relation as a function of redshift, mass and morphology, taking into account also how the details of the cluster internal structure affect the global cluster properties used as mass proxies. And for this we must enrich tSZ observations with high angular resolution, which are among the scientific goals of the NIKA and NIKA2 camera, observing at 150 and 260~GHz.
\begin{figure}
  \centering
    \includegraphics[width=0.22\textwidth]{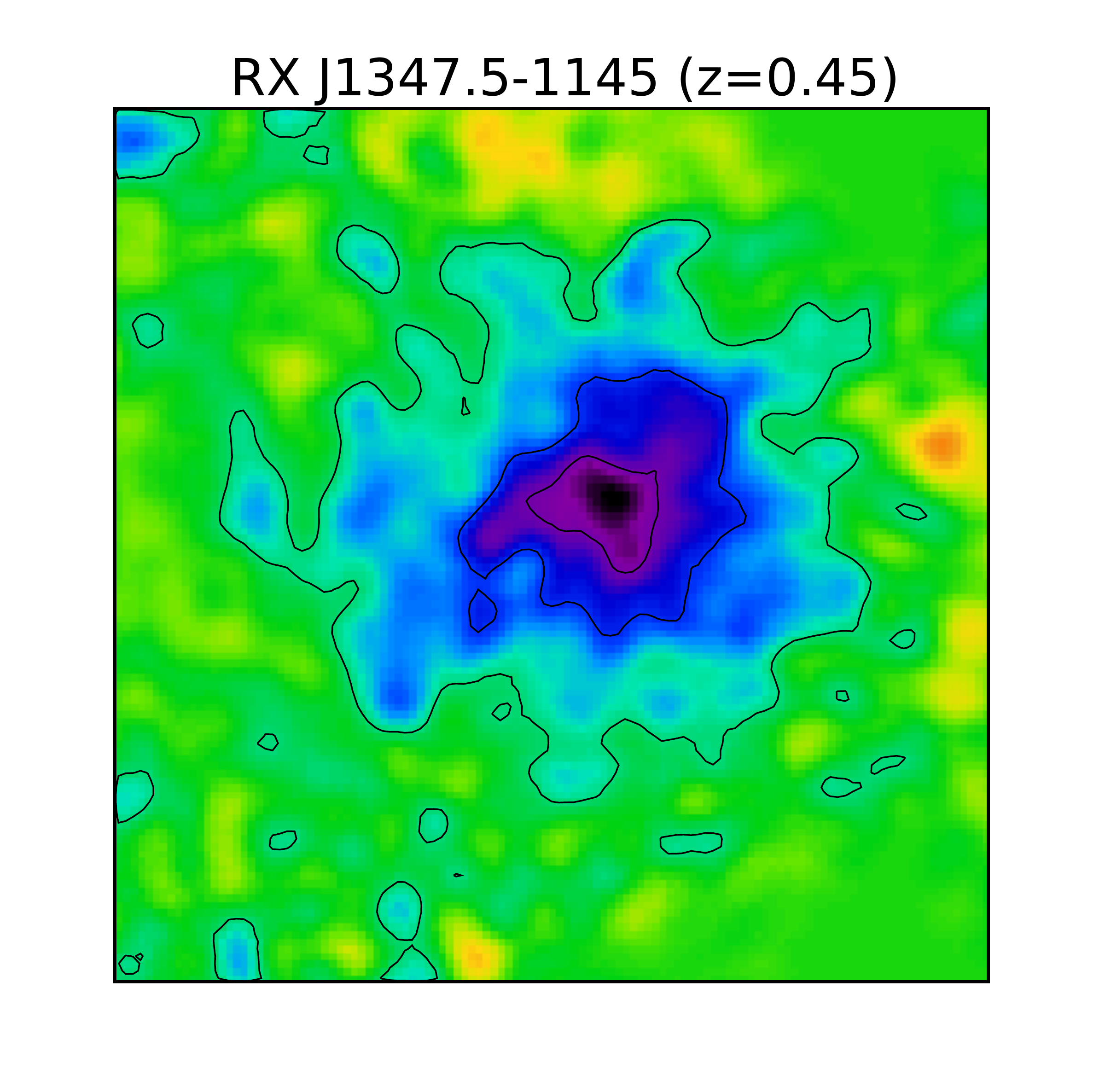}
    \includegraphics[width=0.22\textwidth]{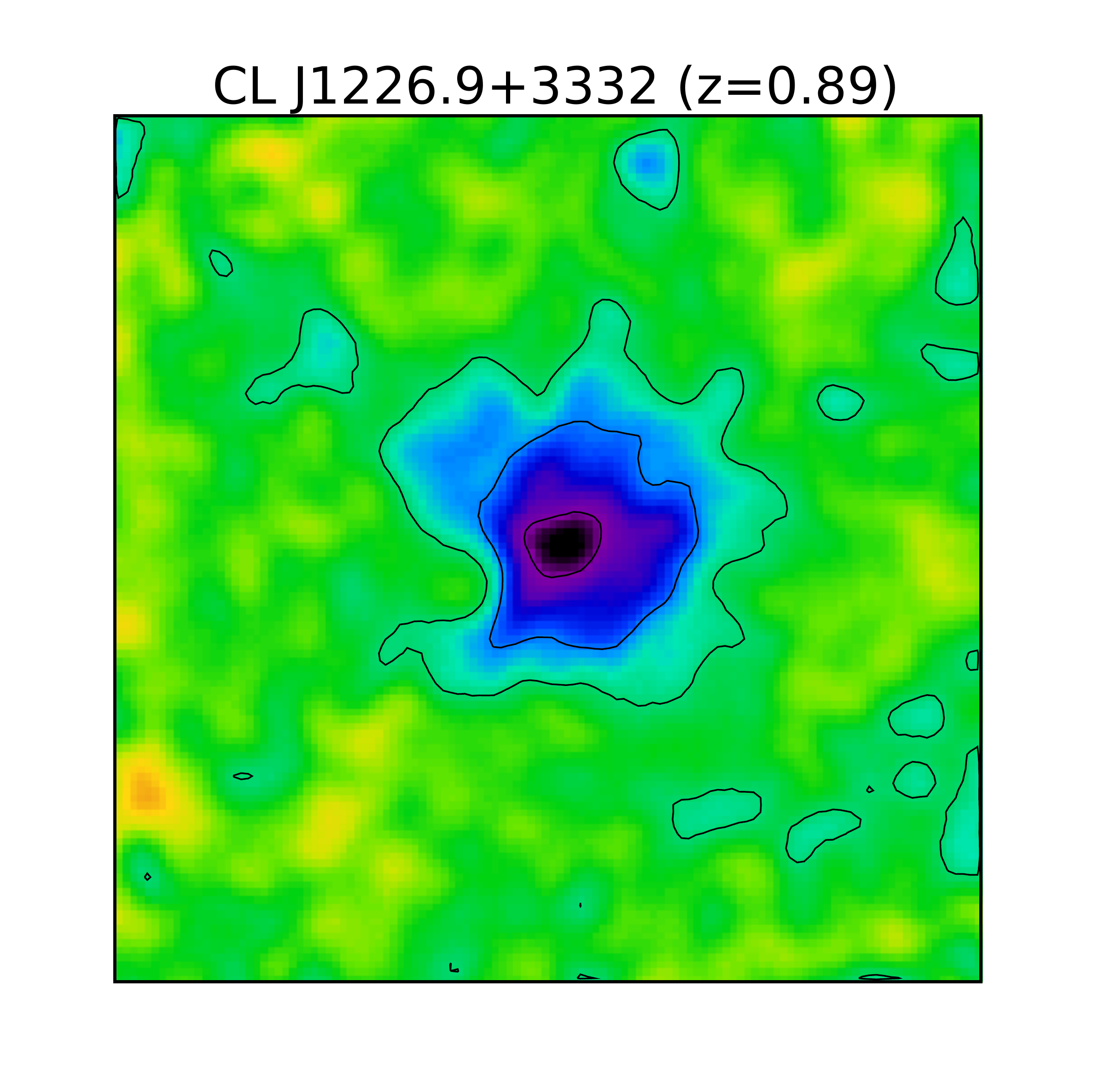}
    \includegraphics[width=0.22\textwidth]{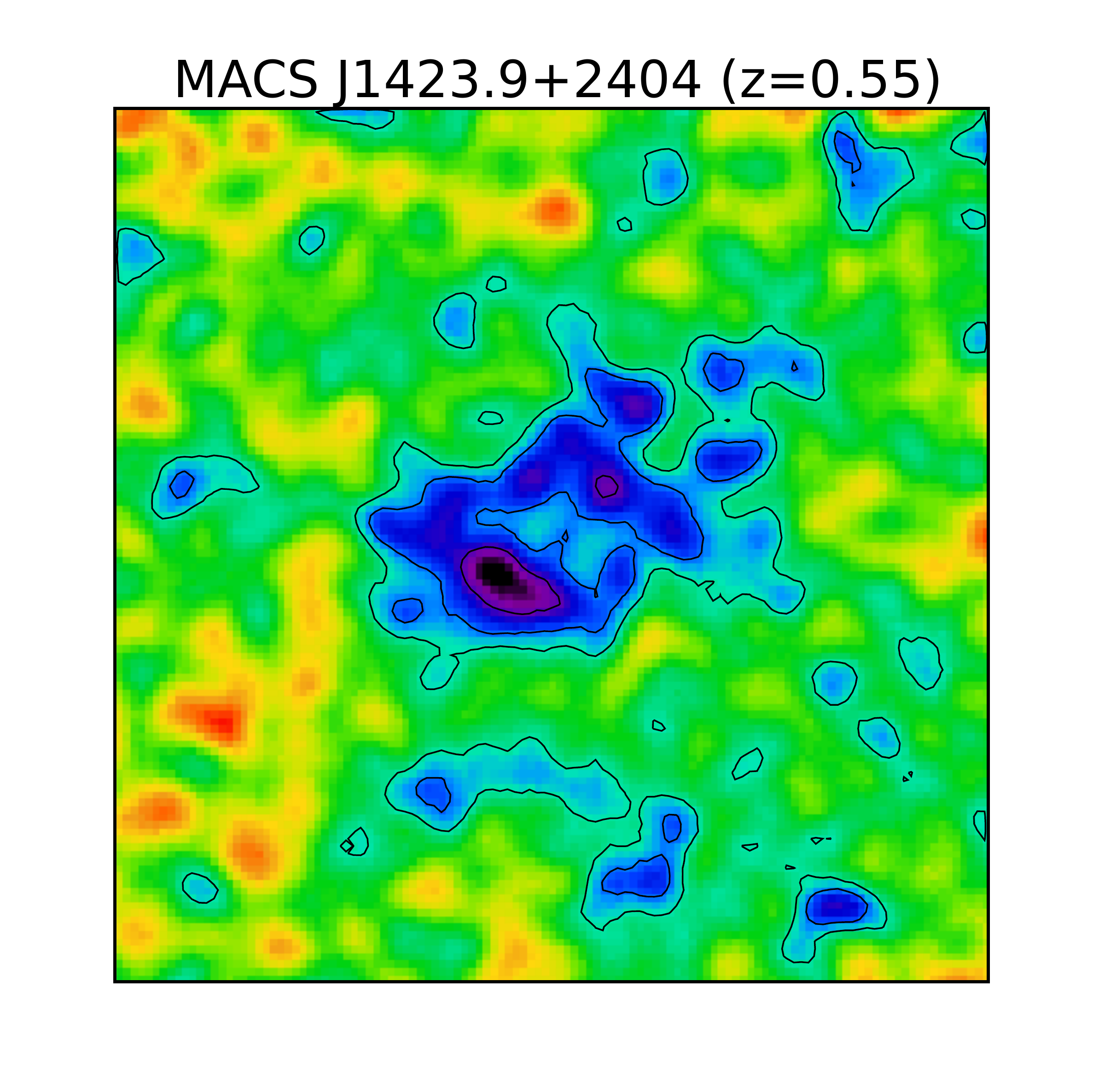}\\
    \includegraphics[width=0.22\textwidth]{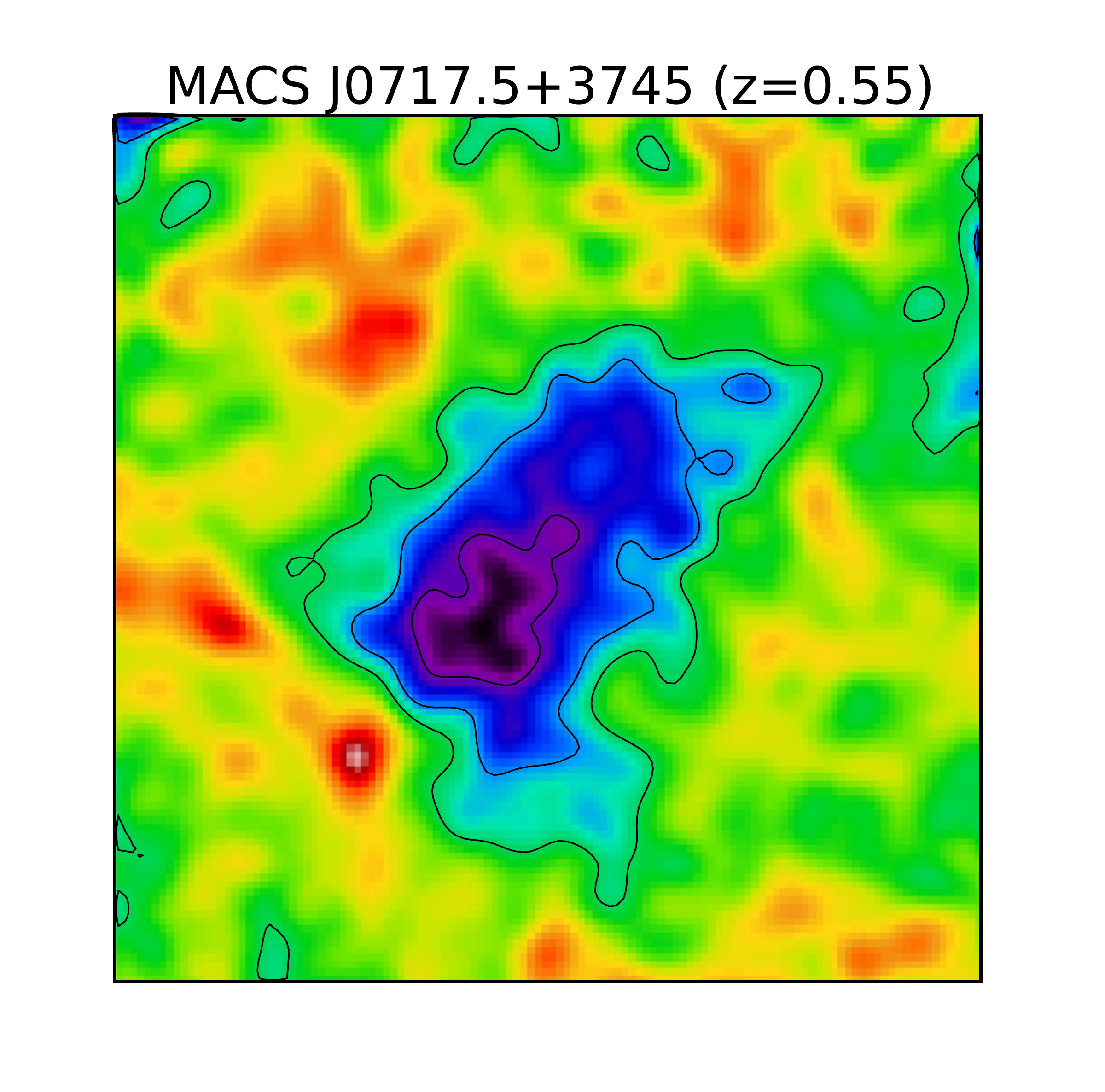}
    \includegraphics[width=0.22\textwidth]{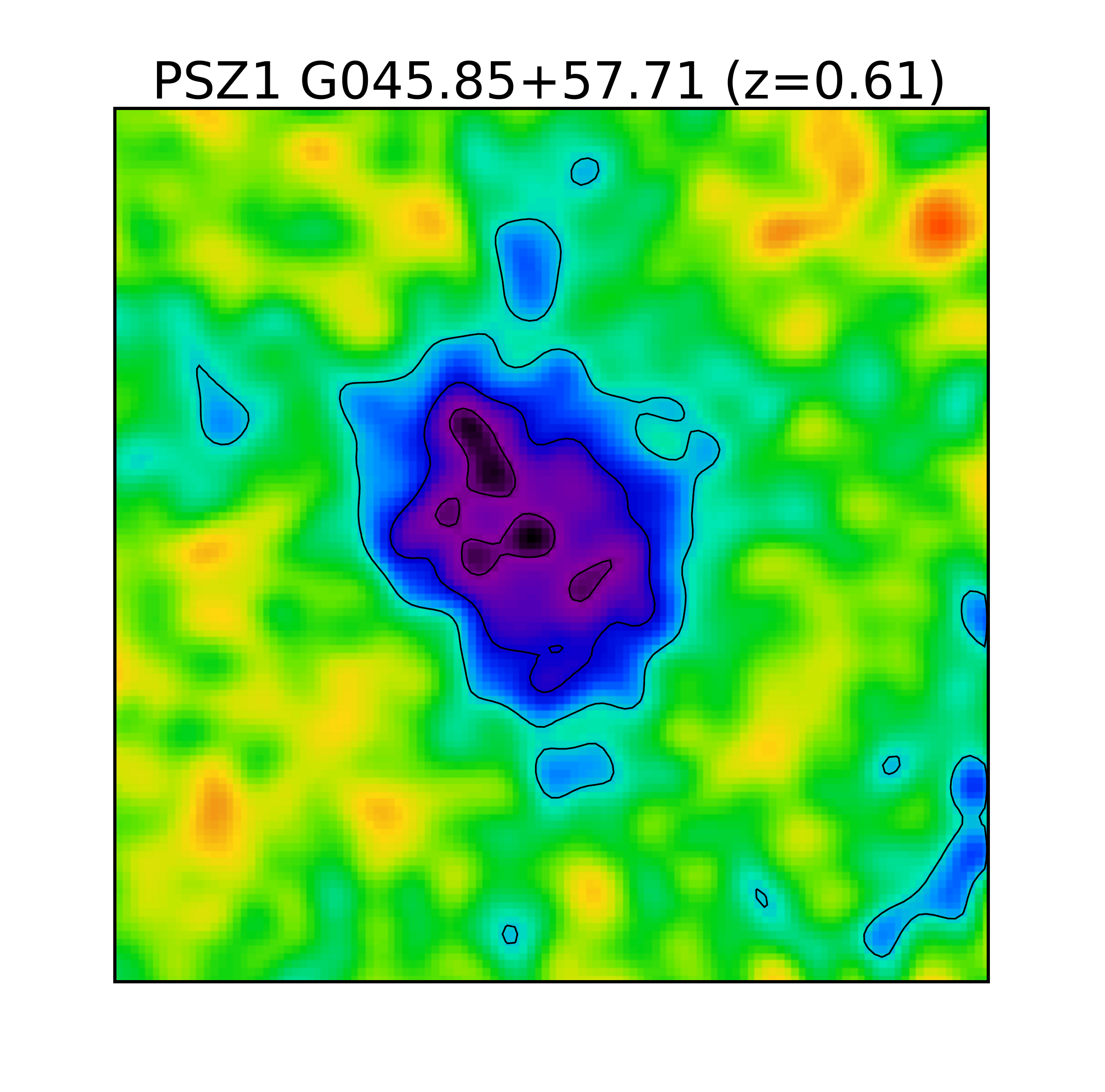}
    \includegraphics[width=0.22\textwidth]{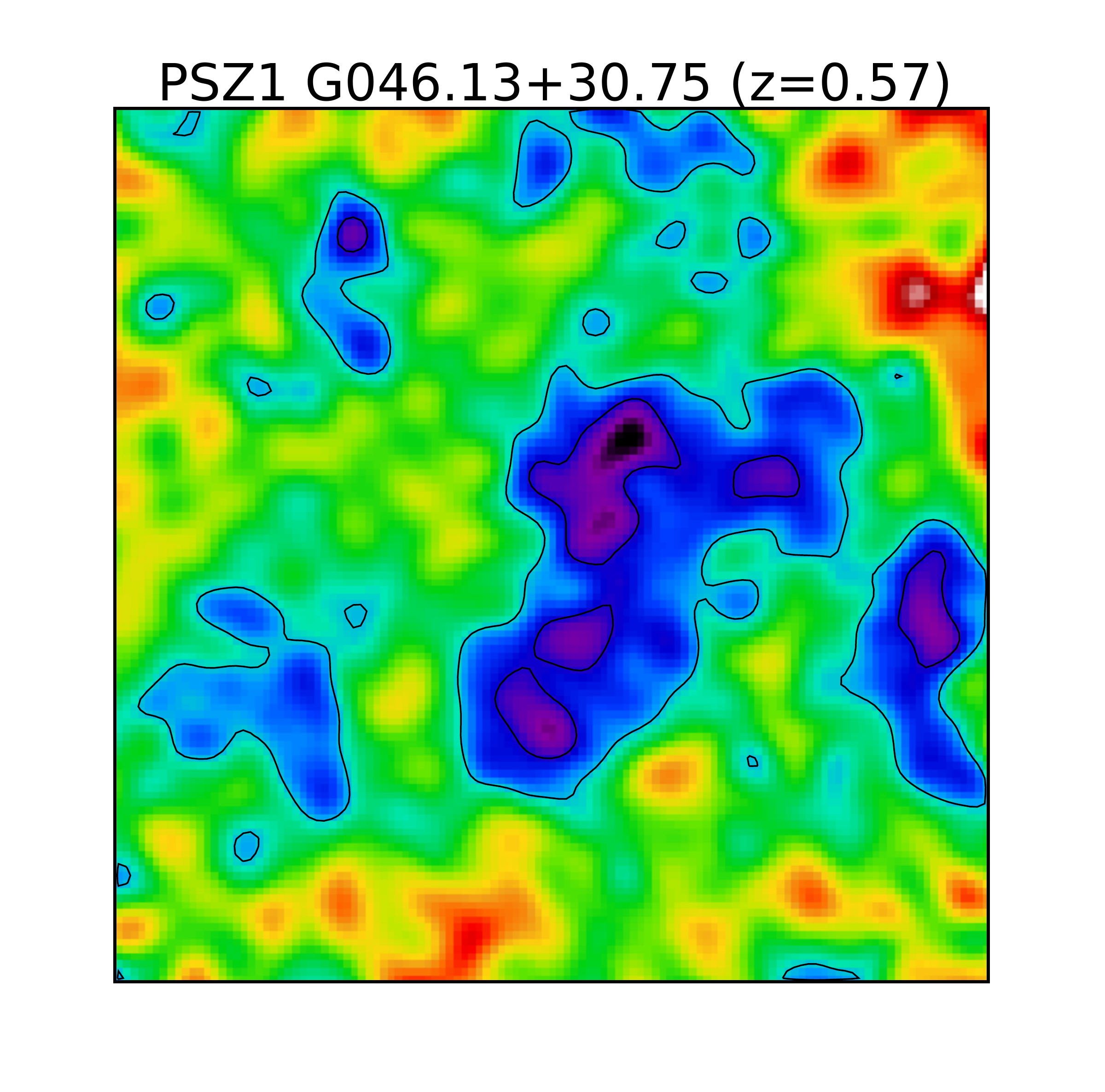}
    \caption{Raw (no point source and foreground removal) 4~arcmin~$\times$~4~arcmin tSZ NIKA maps, obtained at 150~GHz for a sample of six clusters, whose name and redshift are provided in the titles.}
         \label{NIKA_sample}
\end{figure}

\section{From NIKA to NIKA2}\label{NIKA_NIKA2}
The New-IRAM-KIDs-array is a project dedicated to the construction of a dual-band camera customly developed to work at the focal plane of the IRAM 30~m telescope. Located at Pico Veleta, in the Spanish Sierra Nevada, at 2850~m above the sea level, this telescope is one of the two facilities operated by the Institut de Radioastonimie Millimetrique, IRAM, which observes from a site particularly adapted for mm wave astronomy. It is a single dish telescope with a primary mirror with a diameter of 30~m, representing one of the largest and most sensitive to observe at mm wavelength. The size of the primary dish results in a diffraction limited resolution of 17 and 10~arcsec for the 150 and 260~GHz bands respectively. Thus, in order to fully sample the 6.5~arcmin correct Field of View (FoV), a photometric camera for this telescope needs arrays containing hundred of pixels. This is why NIKA uses KIDs, kinetic inductance detectors. Being naturally multi-plexable in the frequency domain, this kind of detectors represents an interesting alternative to the more traditional technologies used at these wavelengths, in the perspective of building arrays with a very large number of detectors. KIDs are printed circuits brought to work at a temperature well below the one at which the material they are made up of become super-conductive. They act like superconductive RLC resonators, characterized by a given resonance frequency. The photons that reach the detector break cooper pairs, and so change the kinetic inductance of the system. This variation in the properties of the circuit results in a change in the resonance frequency, which is propositional to the incoming optical power. This is why, with a single transmission line, we can monitor the resonance frequencies of a large number of pixels, the number being limited only by the bandwidth of the readout electronics.

The project started in 2008, building a pathfinder, NIKA \cite{Monfardini2011}, for the final instrument, NIKA2. NIKA has worked at the focal plane of the 30~m until august 2015. In its final configuration  \cite{Catalano2014}, it counted a total of 356 pixels split over the two bands. The prototype was not able to completely fill the whole FoV of the instrument, but has been a successful experience, used for technical campaigns and also for three observational campaigns open to the scientific community, with proposals from external users. However the most important point is that NIKA has shown state-of-the-art detector performances, which has been a key step for the approval of the NIKA2 project as the camera of choice to be the resident photometric instrument of the IRAM 30~m telescope, for at least the next decade. NIKA2 (New IRAM KID Array 2)\cite{Calvo2016} is a camera based upon kilo-pixel arrays of Kinetic Inductance Detectors. So it fills the entire field of view of the telescope. With a factor ten in the total number of detectors and two arrays at 1.15 mm, allowing us to measure the linear polarization of the incoming light, NIKA2 has been built thanks to the know-how from the experience gained with the prototype. However, it is a completely new instrument, whose development has implied changes at all levels, from the detector arrays themselves all the way up to the optical chain of the 30~m telescope (all the components of the optical chain that follow the secondary mirror have been replaced). The NIKA2 pixels are based on Hilbert type LEKID that can efficiently absorb both polarization, the selection of the direction of the polarisation is done with a wire grid. The NIKA2 cryostat is precooled to $\sim$5~K by two Pulse Tubes (PT) working in parallel. A closed-cycle dilution refrigerator is then used to reach the base working temperature of about 150~mK, which is achieved after 5 days. The system, which is completely cryogen free and can be fully remote controlled, can then be kept cold indefinitely. The multiplexed readout of KID is achieved using dedicated electronics boards called NIKELv15 \cite{Bourrion}. These board can excite and readout up to 400 pixels over a 500~MHz bandwidth.
The camera has been permanently installed at the IRAM 30~m telescope in October 2015. And its commissioning phase is on-going to made accessible to the scientific community at the end of 2016, after a one-year commissioning period. 

\section{NIKA tSZ pilot study}
NIKA2 SZ capabilities have been demonstrated through an SZ pilot study conducted with its pathfinder, NIKA. In the context of this pilot study we have mapped the SZ signal in the direction of six clusters of galaxies (Fig.~\ref{NIKA_sample}), validating the KIDs capabilities when dealing with such a faint and diffuse signal \cite{{Adam2014},{Adam2016}}, even at high redshift \cite{Adam2015}, with complex morphologies \cite{Adam_prep} and at the level of detection of the Planck catalogue of SZ sources \cite{Ruppin_prep}.

\subsection{Well-known cluster -  RX J1347.5-1145}
In November 2012, RX~J1347.5-1145 was identified as the ideal test target. This cluster is one of the most extensively studied, located at intermediate redshift ($z$ = 0.451). It represented a particularly well-suited candidate for the NIKA 5$^{th}$ run, being both compact and strong enough for the prototype FoV and sensitivity. Furthermore, this cluster is a perfect illustration of the complementarity of tSZ effect and X-ray signals: initially thought to be a well-relaxed (cool-core) object according to its first X-ray data, later tSZ observations showed a substructure located at $\sim$ 20~arcsec from the center towards the SE region, interpreted as a hotter, over-pressured component resulting from a merging event. The non-trivial morphology makes this cluster an ideal target to validate NIKA capabilities of probing the details of ICM physics.
The 140 GHz tSZ map, shown in Fig.~\ref{NIKA_sample}, was obtained with a total observing time of 5h~47min and by using the highest frequency band for a dual-band decorrelation of the atmospheric noise \cite{Adam2014}.
In order to extract the signal from the shock, produced by the ongoing merger, we have modeled the relaxed component by considering a generalized Navarro, Frenk and White pressure profile centered at the X-ray position of the system. The model represents well the northern part of the tSZ map while the southern side cannot be explained without including an overpressure component, that is known to be due to the merging of a sub-cluster. This result has allowed us to prove that KIDs arrays are competitive detectors for millimeter wave astronomy, also for the observation of galaxy clusters via the tSZ effect, and that NIKA is able to recover the details of the energy distribution of the gas within clusters, by mapping their tSZ signal.

\begin{figure}
  \centering
    \includegraphics[width=0.4\textwidth]{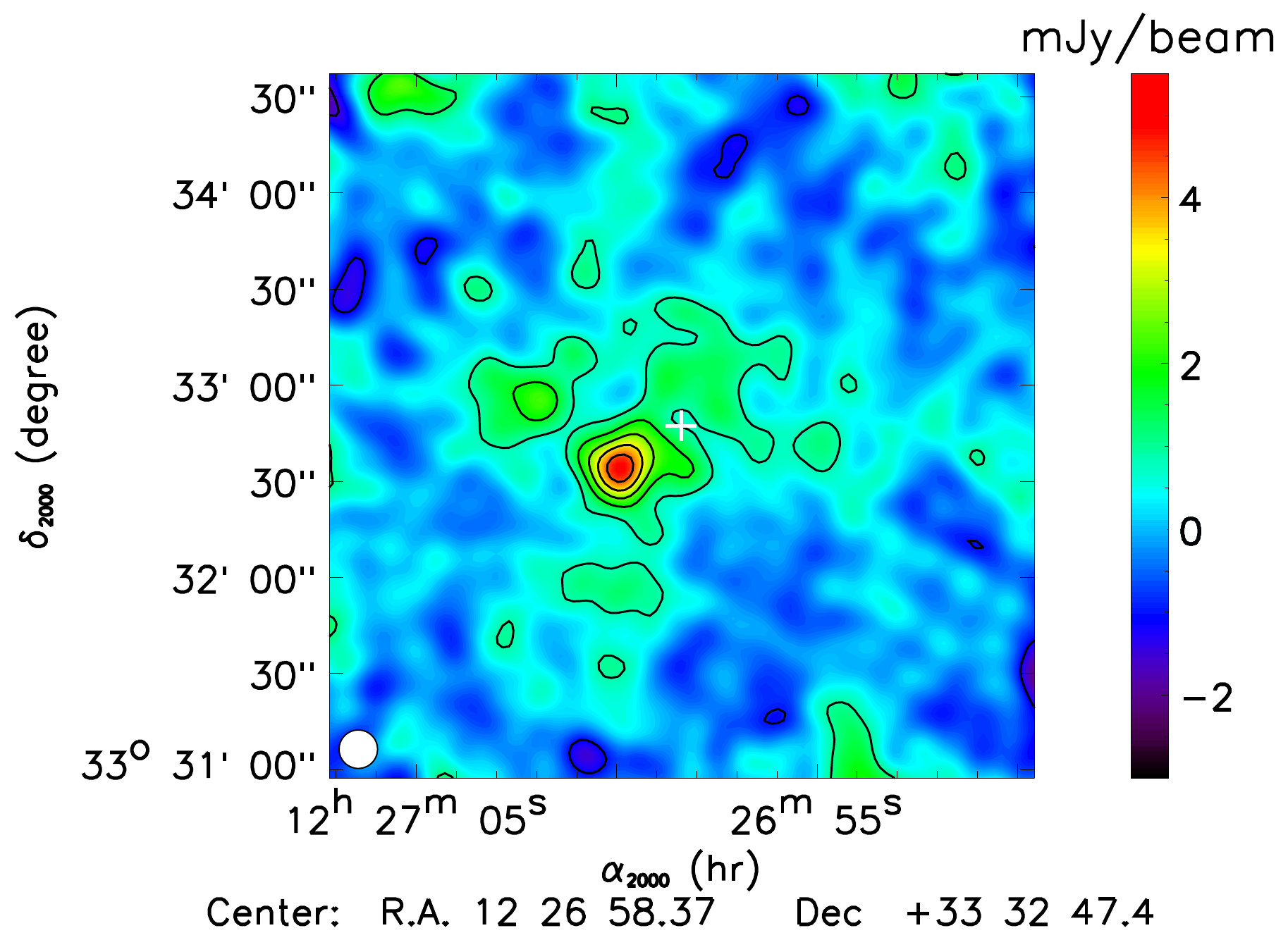}
    \includegraphics[width=0.4\textwidth]{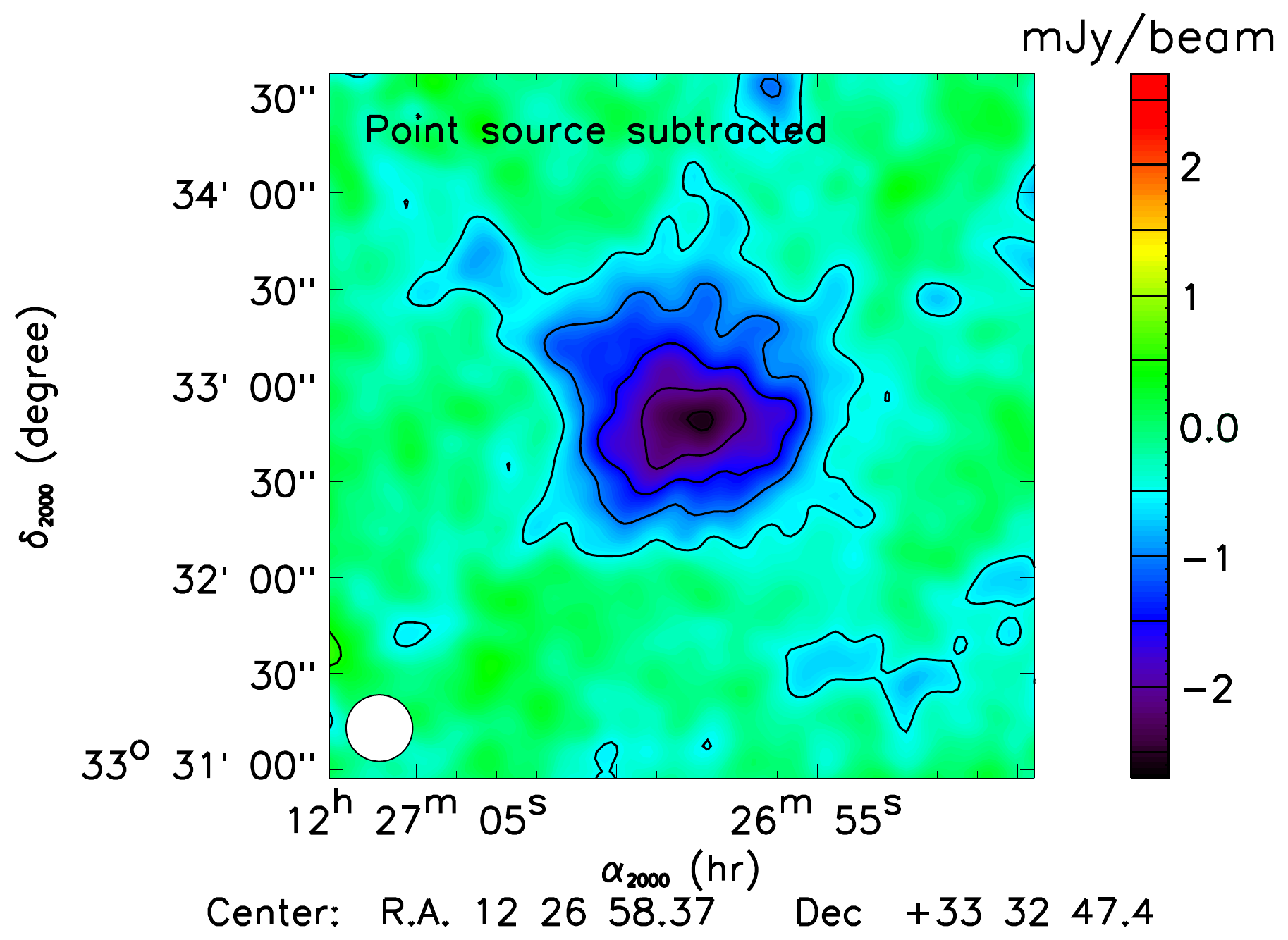}
    \caption{NIKA maps of CL J1226.9+3332 at 260~GHz (left), and at 150~GHz after the subtraction of the contaminating point source (right). Contours are multiples of 3~$\sigma$. The bottom left white circles represent the FWHM of the native beams (12.0 and 18.2 arcsecond), although the displayed images are smoothed with an extra 10 arcsec FWHM Gaussian. The  white cross on the 260 GHz map indicates the position of the X-ray center.}
         \label{CL1227}
\end{figure}

\subsection{High-$z$ cluster - CL J1226.9+3332}
CL~J1226.9+3332 is a hot and massive cluster, located at $z$~=~0.89, that has been chosen to show the tSZ mapping capabilities of NIKA in observing high redshift objects \cite{Adam2015}.
With an overall effective observing time of 7.8 hours, NIKA has provided the first resolved observation of this cluster at these frequencies, as Fig.~\ref{NIKA_sample}~and~~\ref{CL1227} illustrate. The cluster signal is detected in the two bands, with a higher significance at 150~GHz, as expected, since at this frequency the signal is higher. The maps obtained for this cluster point out a further interesting point of the simultaneous dual-band capability. In the case of RX~J1347.5-1145, the highest-frequency channel was used to build a template to remove the atmosphere contamination. This dual-band decorrelation approach has the advantage to enable the recovering the large-scale signal. Given the smaller angular size of CL~J1226.9+3332, a single band common-mode decorrelation is instead used, and the dual band information has provided the detection of a sub-millimeter point source at 260~GHz, 30~arcsec away from the cluster center (Fig.~\ref{CL1227}, left panel). This source also partially compensates the tSZ decrement at 150~GHz (Fig.~\ref{NIKA_sample}). The dual-band capability of NIKA has permitted to account for such contaminant and then to produce a robust reconstruction of the tSZ morphology (Fig.~\ref{CL1227}, right panel). These observations can then be used to reconstruct the pressure profile over a wide range of angular scales (from $\sim$~20 up to $\sim$~200~arcsec) and cluster typical scales (from 0.1~$r_{500}$ up to $r_{500}$, $r_{500}$ being the radius at which the cluster mean density is equal to 500 times the critical density of the Universe, at the cluster redshift).
The reconstructed tSZ map of CL~J1226.9+3332 has been used to constrain its pressure distribution, as well as the thermodynamics of the ICM gas by combining it with X-ray data.

\subsection{Relaxed cluster - MACS~J1423.8+2404}
\begin{figure}
  \centering
   \includegraphics[width=0.37\textwidth]{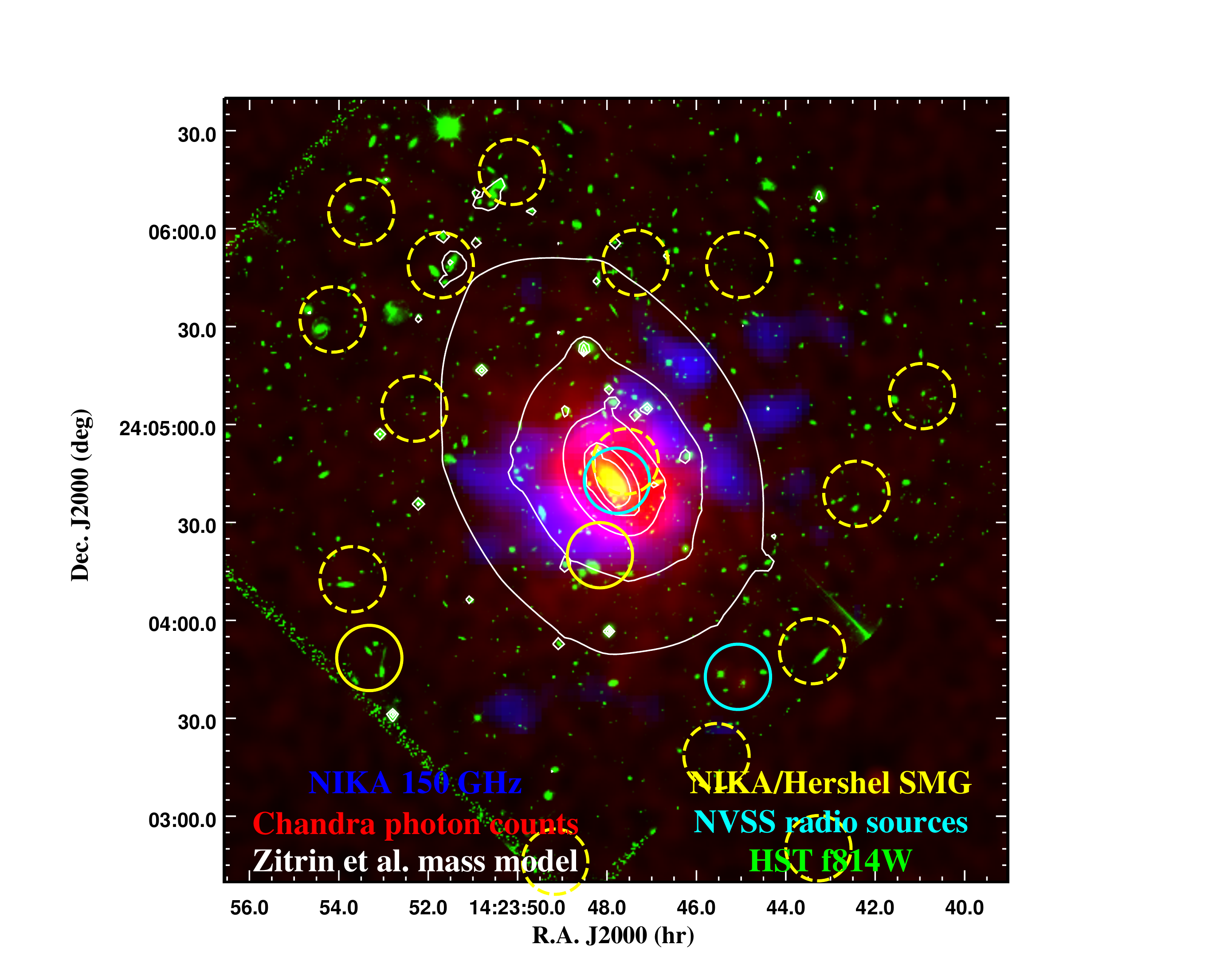}
    \includegraphics[width=0.4\textwidth]{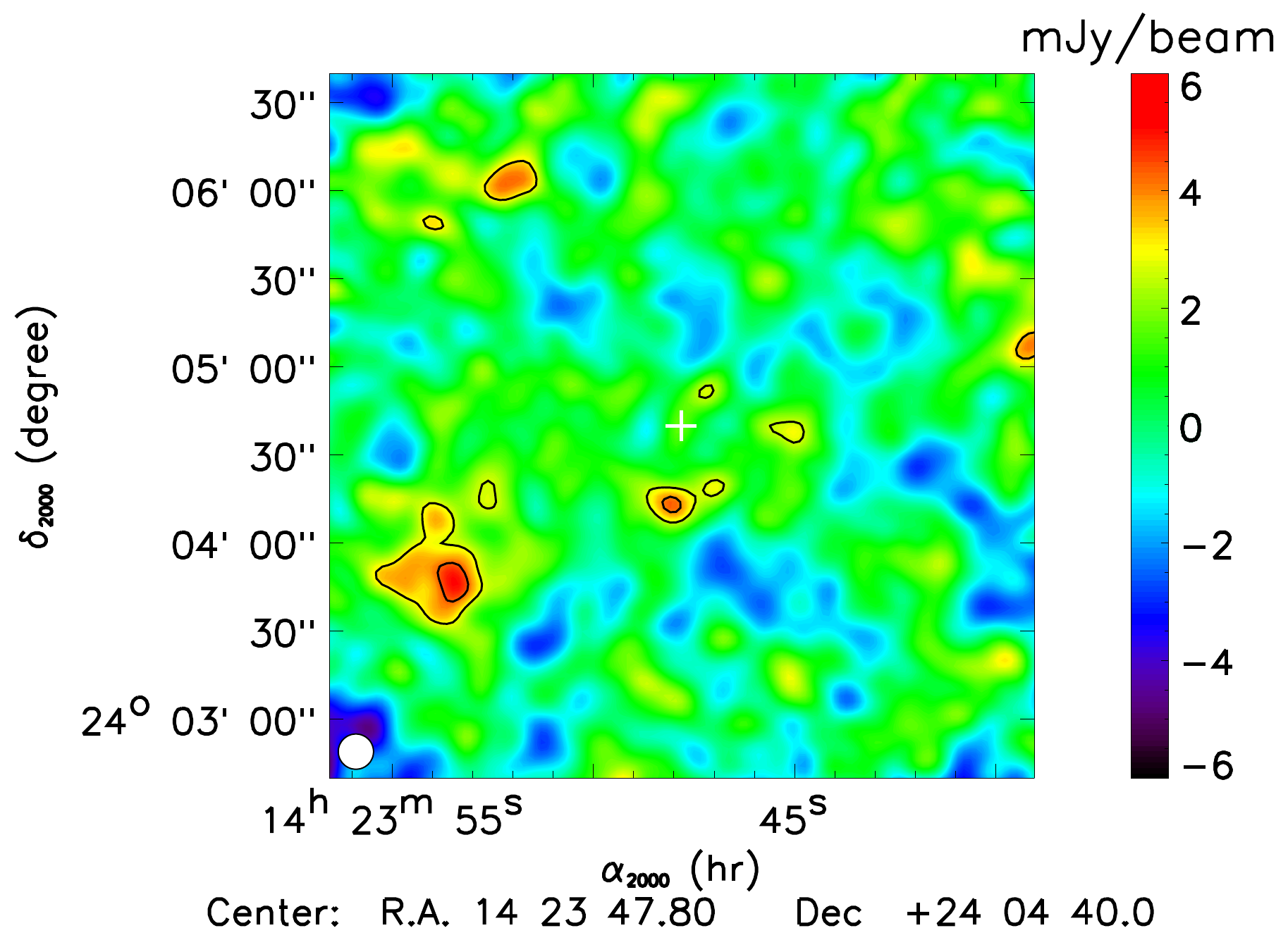}
    \includegraphics[width=0.4\textwidth]{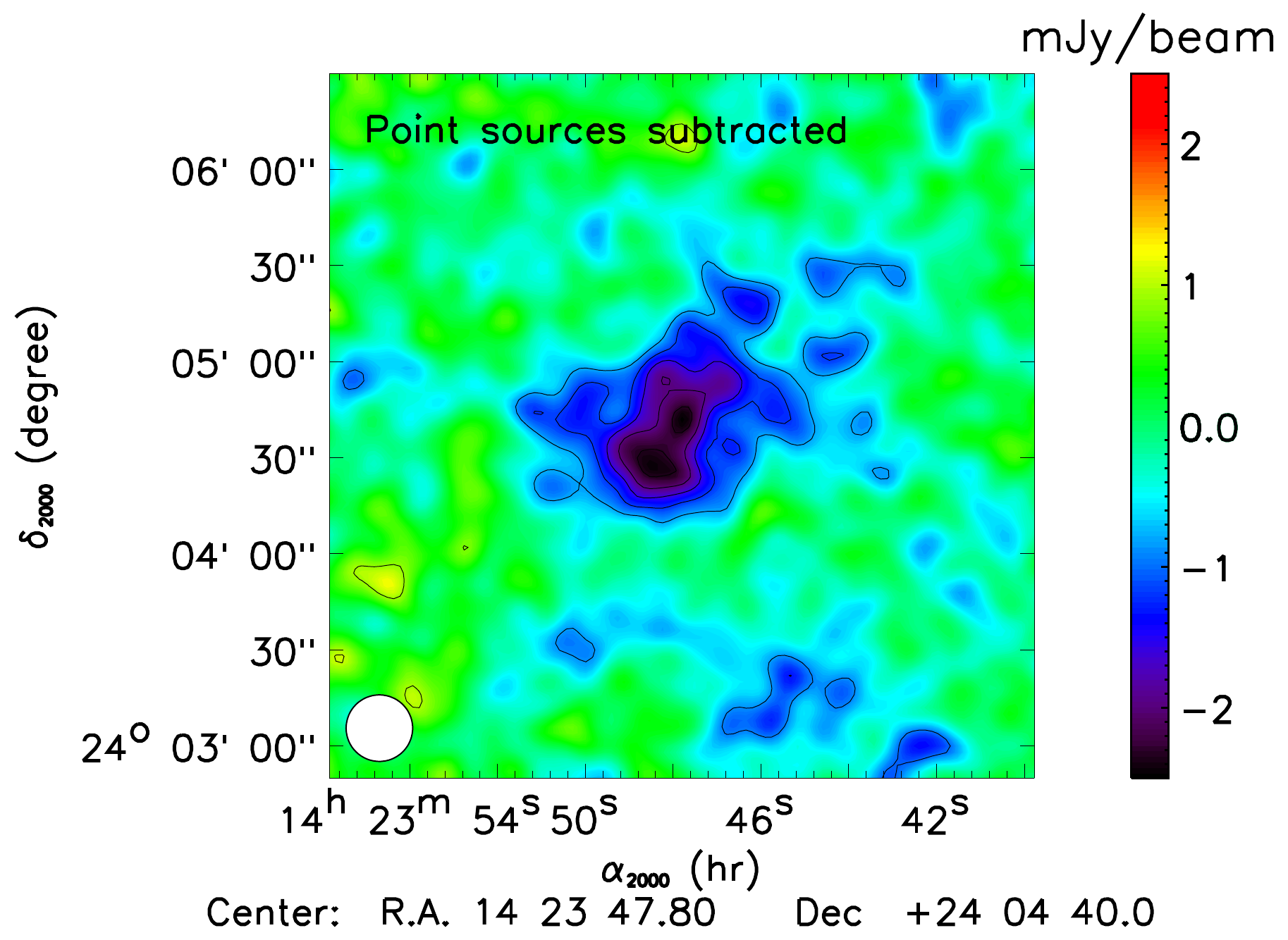}
    \includegraphics[width=0.4\textwidth]{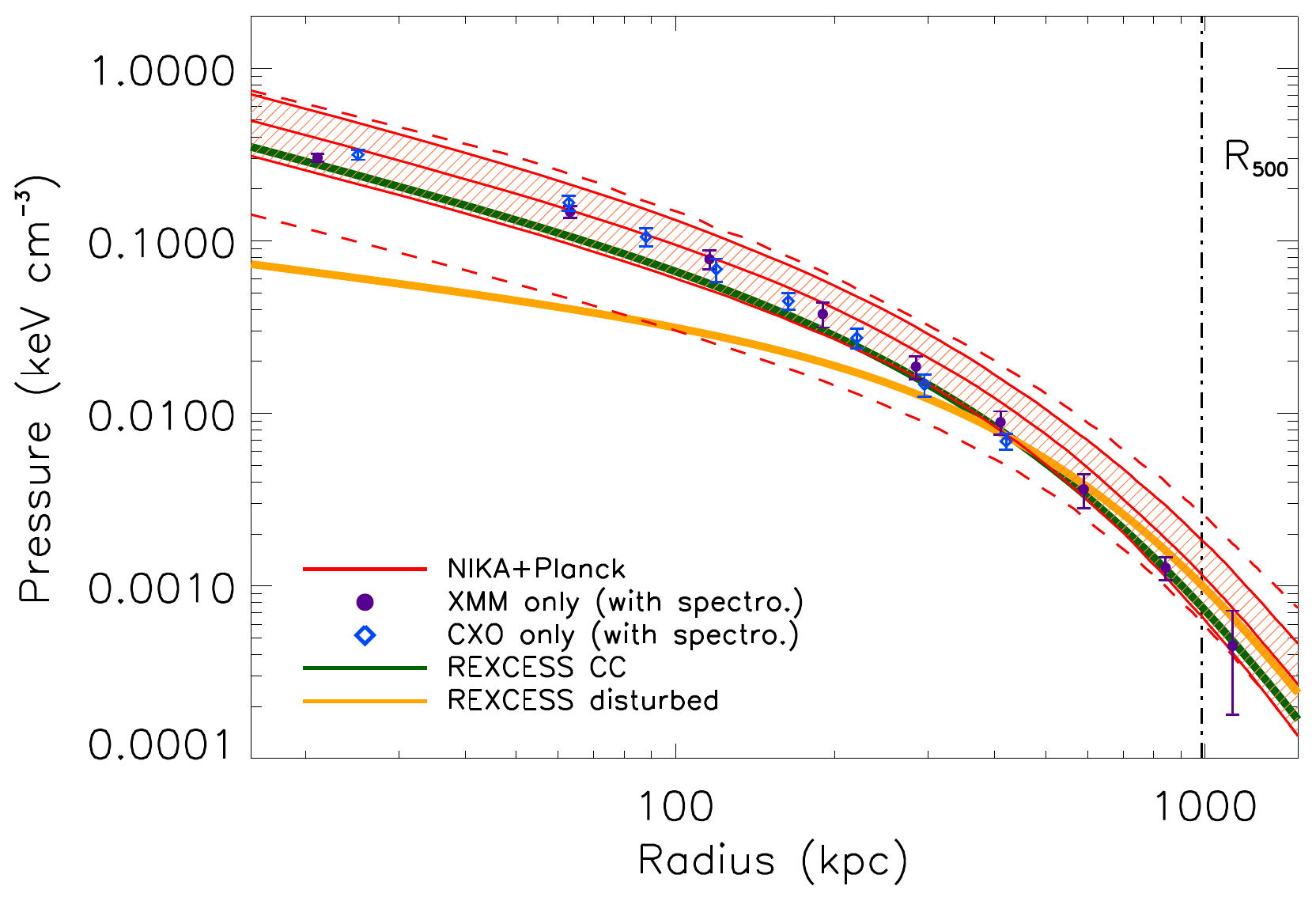}
    \caption{{\bf Upper left:} Composite multi-wavelength image of MACS~J1423.8+2404. In blue the NIKA 150~GHz map (tSZ signal). In red the Chandra photon counts (tracing the electronic density). In green the Hubble Space Telescope data. The surface mass distribution model is shown by the white contours while the yellow circles correspond to millimeter sources locations obtained with the NIKA 260~GHz channel (solid line) and identified using {\it Herschel} (dashed-line). The cyan circles show the radio point sources present in the field (VLA). {\bf Upper right:} NIKA 260~GHz map, the white cross indicates the X-ray center. {\bf Bottom left:} NIKA 150~GHz map after point source subtraction. {\bf Bottom right:} In red the NIKA constraints on the deprojected pressure radial profile of MACS~J1423.8+2404. The Chandra and XMM-Newton only measurements are respectively shown with blue diamonds and purple dots. The green and orange solid lines respectively allows also comparison with the universal cluster pressure profiles of cool core and morphologically disturbed clusters, as obtained with REXCESS, a representative sample of nearby X-ray clusters.}
         \label{MACS1424}
\end{figure}

MACS~J1423.8+2404 has been chosen to test the observation of relaxed clusters as well as the impact of point sources contamination on the reconstruction of the pressure distribution with NIKA\cite{Adam2016}. One of the main challenges in observing relaxed clusters is that they often host a strong central radio source, which can significantly compensate the tSZ decrement. Furthermore, in the context of high angular resolution tSZ observations, in general the removal of the contamination from point sources will be one of the main challenges. Galaxy clusters are in fact crowded environment, containing galaxies that can host radio sources or a significant amount of dust. Finally, clusters at intermediate redshift provide optimal lenses, which can magnify sub-millimeter background galaxies, in addition to the possible presence of foreground contaminating sources.
In Fig.~\ref{MACS1424} we show a multi-wavelength view of MACS J1423.8+2404, as well as the NIKA maps at 150 and 260~GHz. In the 4 $\times$ 4 arcmin$^2$ field around MACS~J1423.8+2404, 19 point sources have been identified and ancillary sub-millimeter ({\it Herschel}) and radio (SZA, OVRO/BIMA, VLA and NVSS). The target was detected by NIKA at 150~GHz (4.5~$\sigma$) in only 1.47 hours and the ring-like shape at 150~GHz (Fig. \ref{NIKA_sample}) shows evidence for the presence of a contaminating central radio source, as expected. But this cluster also hosts sub-millimeter point sources, two of which are also detected on the NIKA 260~GHz map (Fig. \ref{MACS1424}, upper right panel). Previous radio observations have been used to extrapolate the fluxes to the NIKA bands, while sub-millimeter sources have been extrapolated by modeling their SED with a gray-body spectrum, obtained complementing the NIKA data with those from the {\it Herschel} satellite.
The bottom left panel of Fig.~\ref{MACS1424} shows the 150~GHz map after point source subtraction. The results obtained for this cluster allow us to show that, if external data at complementary wavelengths are available, point sources contamination can be estimated and subtracted from the NIKA tSZ map. Nevertheless, the presence of a point source at the cluster center, without strong priors on its flux, can severely limit our ability to constrain the inner pressure profile and the cluster morphology. However, the resolved nature of the NIKA observations allows the large-scale pressure distribution to be almost unaffected by the presence of point source. 
The NIKA tSZ data have been combined also with the Planck ones, as well as with the X-ray observations obtained both with Chandra and XMM-Newton
The comparison with X-ray only data shows good agreement for the pressure (Fig.~\ref{MACS1424}, bottom right panel), temperature, and entropy profiles, all indicating that MACS J1423.8+2404 is a dynamically relaxed cool-core system. The tSZ-X joint analysis conducted on this cluster has also shown that the quality of the NIKA tSZ mapping enables to finally use the tSZ and X-ray complementarity to constrain the temperature and entropy profiles of galaxy clusters independently from X-ray spectroscopy, using density and pressure profiles.
Moving to high redshift, X-ray observations become time expensive and high-quality X-ray mapping becomes challenging because of redshift dimming. In this context, the results we have obtained on this cluster show that we are able to constrain the ICM thermodynamics with a good accuracy by combining resolved NIKA tSZ observations and X-ray mapping.

\section{NIKA2 tSZ large program}
On the basis of the success of the NIKA tSZ pilot study, 300~hours of the NIKA2 telescope guaranteed time have been assigned to develop an {\it SZ Large Program}. The main objective of this program is to obtain high resolution tSZ observations for a representative sample of clusters at intermediate and high redshift (z $>$ 0.5), to study the evolution of the cluster physical properties across cosmic times. At present, cluster derived cosmological constraints are in fact limited by our understanding of the impact of the details of cluster astrophysics and this kind of study is then mandatory to make clusters a competitive probe for the future challenges of precision cosmology. NIKA2 (Sect. \ref{NIKA_NIKA2}) is well adapted for high angular resolution follow-up observations of SZ clusters, because of its large number of high sensitive detectors observing at two frequency bands (150~and 260~GHz), its large field of view (6.5~arcmin) and the resolution allowed by a 30~m telescope. 

With this camera, we intend to observe a large ($\sim$ 50) and cosmologically representative sample of clusters of galaxies, with redshift $\geq$ 0.5. The main output of the program will be the study of the redshift evolution of the cluster pressure profiles and of the scaling laws relating cluster direct observables (i.e. the integrated Compton parameter, Y) to their total masses (M$_{tot}$). This can be achieved with NIKA2 tSZ observations and by combining them with ancillary data, including X-rays and optical observations, leading to significant improvements on the use of clusters of galaxies to draw cosmological constraints.
Our target selection strategy is mainly driven by the need of an homogeneous coverage in SZ flux. Especially when dealing with SZ, a flux-selected subset of the cluster population can be considered as representative of a sample and not biased towards a given morphology. Such a sample is then adapted to the goals of our large program: derive relations that can be applicable to the whole cluster population,  achieving a good global characterization of the cluster population and an improved control of systematics due to cluster astrophysics, up to high-$z$.
Our selection criterion follows in fact the approach adopted to build the REXCESS \cite{REXCESS} sample, an XMM-Newton large program dedicated to the in-depth study of a representative sample of 33 clusters (0.055 $<$ z $<$ 0.183) that has been used to build the universal pressure profile for the ICM \cite{Arnaud2010}. 

In order to fulfill this goal within the total available observing time, 300 hours, we have considered the following target selection criteria:
\begin{itemize}
  \item [-] clusters belonging to SZ selected samples (already existing tSZ based cluster samples from Planck and ACT) for which we already have the redshift information and estimates of the total tSZ flux (to add a further constrain at large angular scale);
  \item [-] z $>$ 0.5, to which the NIKA2 FoV is the most adapted;
  \item [-] dec $>$ -11, to ensure observability of the sources from the Pico Veleta site.
\end{itemize}
The overall observation time has been distributed among the different clusters of the sample taking into account the need of having a homogeneous quality of the maps, at a given characteristic radius, for the whole sample.

\section{Conclusions}
At present, cluster derived cosmological constraints are limited by our understanding of the impact of the details of cluster astrophysics.
The development of precision cosmology with clusters requires high angular resolution and multi-probe studies (X-ray, tSZ, lensing, optical) up to high redshift. On the SZ side what must be done is to continue enriching catalogs up to high $z$, but also to calibrate the properties of the cluster population at the different epochs: without these tools in our hands we cannot fully exploit catalogues, and we will remain limited by astrophysical systematics.

With NIKA2, a kilo-pixel KIDs based camera installed at the focal plane of the IRAM 30~m telescope, we will produce high-angular resolution ($\sim$~20~arcsec) tSZ maps of a representative sample of clusters, belonging SZ selected catalogues and located at $z >$~0.5. The NIKA2 tSZ capabilities have been demonstrated through a pilot study, conducted with its pathfinder, NIKA. Between 2012 and 2014 NIKA has observed six clusters, providing excellent results up to high redshift ($z \sim$~0.9) and at the level of detection of the Planck catalogue of SZ sources.

The NIKA2 tSZ Large Program will spend 300~hours of telescope time observing a statistically significant ($\sim$~50), representative sample of clusters of galaxies at intermediate and high redshift
The angular resolution of the system will enable a detailed examination of cluster morphology through the tSZ effect. This will enable the study of the calibration of the SZ flux as a mass proxy, its evolution with redshift and cluster dynamics, the redshift evolution of the universal cluster pressure profiles, as well as deviations from its mean behavior due to cluster complex astrophysics and thermodynamical history.
This kind of study is mandatory to achieve cluster derived precision cosmology.

\section*{Acknowledgments}
We would like to thank the IRAM staff for their support during the campaigns. 
The NIKA dilution cryostat has been designed and built at the Institut N\'eel. 
In particular, we acknowledge the crucial contribution of the Cryogenics Group, and 
in particular Gregory Garde, Henri Rodenas, Jean Paul Leggeri, Philippe Camus. 
This work has been partially funded by the Foundation Nanoscience Grenoble, the LabEx FOCUS ANR-11-LABX-0013 and 
the ANR under the contracts "MKIDS", "NIKA" and ANR-15-CE31-0017. 
This work has benefited from the support of the European Research Council Advanced Grant ORISTARS 
under the European Union's Seventh Framework Programme (Grant Agreement no. 291294).
We acknowledge fundings from the ENIGMASS French LabEx (R. A., F. R. and B. C.), 
the CNES post-doctoral fellowship program (R. A.),  the CNES doctoral fellowship program (A. R.) and 
the FOCUS French LabEx doctoral fellowship program (A. R.).

\end{document}